\documentclass[prd,aps,showpacs,tightenlines,twocolumn,nofootinbib]{revtex4}
\usepackage{graphicx}

\newcommand{\CL}{{\cal L}}

\newcommand{\bear}{\begin{array}}  \newcommand{\eear}{\end{array}}
\newcommand{\bea}{\begin{eqnarray}}  \newcommand{\eea}{\end{eqnarray}}
\newcommand{\beq}{\begin{equation}}  \newcommand{\eeq}{\end{equation}}
\newcommand{\bef}{\begin{figure}}  \newcommand{\eef}{\end{figure}}
\newcommand{\bec}{\begin{center}}  \newcommand{\eec}{\end{center}}
\newcommand{\non}{\nonumber}  
  
\newcommand{\lkk}{\left[}  \newcommand{\rkk}{\right]}
  
\newcommand{\del}{\partial}  

\newcommand{\bib}{\bibitem} 
\newcommand{\la}{\left\langle} \newcommand{\ra}{\right\rangle}

\def\AAA#1#2#3{Astron. Astrophys. {\bf #1}, #2 (20#3)}

\def\APJJ#1#2#3{Astrophys. J. {\bf #1}, #2 (20#3)}

\def\APJSS#1#2#3{Astrophys. J. Suppl. {\bf #1}, #2 (20#3)}

\def\JCAPP#1#2#3{JCAP {\bf #1}, #2 (20#3)}

\def\NPB#1#2#3{Nucl. Phys. {\bf B#1}, #2 (19#3)}
\def\NPBB#1#2#3{Nucl. Phys. {\bf B#1}, #2 (20#3)}

\def\PLB#1#2#3{Phys. Lett. B {\bf #1}, #2 (19#3)}
\def\PLBB#1#2#3{Phys. Lett. B {\bf #1}, #2 (20#3)}
\def\PLBold#1#2#3{Phys. Lett. {\bf#1B}, #2 (19#3)}

\def\PRD#1#2#3{Phys. Rev. D {\bf #1}, #2 (19#3)}
\def\PRDD#1#2#3{Phys. Rev. D {\bf #1}, #2 (20#3)}

\def\PRL#1#2#3{Phys. Rev. Lett. {\bf#1}, #2 (19#3)}

\def\PRTT#1#2#3{Phys. Rep. {\bf#1}, #2 (20#3)}
\def\PTP#1#2#3{Prog. Theor. Phys. {\bf #1}, #2 (19#3)}

\def\SJNP#1#2#3{Sov. J. Nucl. Phys. {\bf #1}, #2 (19#3)}

\newcommand{\lesssim}{ \mathop{}_{\textstyle \sim}^{\textstyle <} }

\begin{document}

\title{Spontaneous baryogenesis in flat directions}
\author{Fuminobu Takahashi}
\affiliation{Research Center for the Early Universe, University of
Tokyo, Tokyo 113-0033, Japan}
\author{Masahide Yamaguchi}
\affiliation{Physics Department, Brown University, Providence, RI 02912,
USA}
\date{\today}
\begin{abstract}
  { We discuss a spontaneous baryogenesis mechanism in flat
    directions. After identifying the Nambu-Goldstone mode which
    derivatively couples to the associated $U$(1) current and rotates
    due to the A-term, we show that spontaneous baryogenesis can be
    naturally realized in the context of the flat direction. As
    applications, we discuss two scenarios of baryogenesis in
    detail. One is baryogenesis in a flat direction with a vanishing $B-L$
    charge, especially, with neither baryon nor lepton charge, which was
    recently proposed by Chiba and the present authors. The other is a
    baryogenesis scenario compatible with a large lepton asymmetry.}
\end{abstract}

\pacs{98.80.Cq \hspace{5.6cm} RESCEU-47/03, BROWN-HET-1376} \maketitle


\section{Introduction}
\label{sec:introduction} 

Baryon asymmetry is a great mystery in cosmology and particle physics.
The recent observations of the cosmic microwave background anisotropy
by the Wilkinson Microwave Anisotropy Probe (WMAP) show that the
baryon to entropy ratio is $n_{B} / s \sim 9 \times 10^{-11}$
\cite{WMAP}, which roughly coincides with the value inferred from big
bang nucleosynthesis (BBN)~\cite{BBN}. Although many scenarios have been
proposed so far to explain such an asymmetry, a spontaneous
baryogenesis mechanism proposed by Cohen and Kaplan is special in that
it works even in thermal equilibrium \cite{CK}. The other scenarios
require a deviation from thermal equilibrium, which often imposes
severe constraints on the scenario.

The supersymmetric  theory is one of the most attractive
extensions of standard model particle physics because it stabilizes
the electroweak scale against radiative corrections and realizes
the unification of standard gauge couplings. In the supersymmetric
theory, flat directions are ubiquitous and their existence distinguishes
supersymmetric theories from ordinary ones \cite{EM}.  Although there are
no classical potentials along flat directions in the supersymmetric
limit, these directions can be lifted by both the supersymmetry-breaking
effects and the nonrenormalizable operators with some cutoff scale. In
particular, during inflation, a flat direction receives a mass squared
proportional to the Hubble parameter squared, so that the flat direction
acquires a large expectation value for the negative mass squared. As
inflation ends and the Hubble parameter becomes small enough, the flat
direction starts to oscillate and rotate due to the so-called A-term.

Associated with such dynamics of the flat direction, there are two
different sources of baryon and/or lepton asymmetries. First, in the
case that the flat direction carries the baryon and/or lepton numbers,
the rotation due to the A-term implies that baryon and/or lepton
asymmetries are generated as a condensate of the flat direction. After
the decay of the flat direction, such asymmetries are released to the
ordinary quarks and leptons. This is the so-called Affleck-Dine (AD)
mechanism \cite{AD}. Another source is the coupling between the phase of
the flat direction and the baryon and/or lepton current. In fact, as
shown later, the phase of the flat direction couples to such a current
derivatively. Then, the rotation of the flat direction due to the A-term
leads to CPT violation, so that baryon and/or lepton asymmetries are
generated for light particles if the current violating interactions are
in good thermal equilibrium. This is actually the realization of the
spontaneous baryogenesis proposed by Cohen and Kaplan.

Until now, in almost all research, only the first source has been
considered. However, the AD mechanism applies only to flat directions
with nonzero $B-L$ charge because otherwise sphaleron effects wash
out the produced baryon asymmetry. Recently, it was shown that, if
$Q$-balls are formed, the AD mechanism can be applied to flat
directions with vanishing $B-L$ charge \cite{Kusenko, Enqvist,
  Kasuya1}. This is because $Q$-balls can protect the $B+L$ asymmetry
from the sphaleron effects. However, the AD mechanism does not work at
all for flat directions with neither baryon nor lepton charge, that
is, $B=L=0$.

On the other hand, very recently, Chiba and the present authors showed
that baryogenesis is possible even for such a flat direction in the
context of a second source if the flat direction has another charge
\cite{CTY}. In this paper, we investigate the second source in detail,
that is, a spontaneous baryogenesis mechanism in flat directions.
Spontaneous baryogenesis in another context is considered in
\cite{LFZ,FNT,Yamaguchi}. In Sec. II, we identify the Nambu-Goldstone
(NG) mode which derivatively couples to the associated U(1) current and
rotates due to the A-term, and we give a general discussion on
spontaneous baryogenesis in flat directions. In Sec III, as
applications, we concentrate on two scenarios of baryogenesis in
particular. One is baryogenesis in a flat direction with vanishing $B-L$
charge, especially, with neither baryon nor lepton charge. As listed in
Ref. \cite{tony}, in fact, there are many flat directions in which the $B-L$
charge vanishes. The other is a baryogenesis scenario compatible with a
large lepton asymmetry. The baryon-to-entropy ratio inferred from recent
results of WMAP roughly coincides with that inferred from BBN. However,
according to a detailed analysis \cite{BBN2}, the best fit value of
the effective number of neutrino species $N_{\nu}$ is significantly
smaller than $3.0$. Of course, $N_{\nu} = 3.0$ is consistent at $\sim 2
\sigma$, and such discordance may be completely removed as observations
are further improved and the errors are reduced. However it is probable
that such small discrepancies are genuine and suggest additional physics
in BBN and the CMB. One interesting possibility eliminating such
discrepancies is the presence of a large and positive lepton asymmetry
of electron type \cite{KS}, so we discuss such a possibility in this
context. Finally, we devote Sec. IV to discussion and conclusions.

\section{Spontaneous Baryogenesis in Flat Direction}
\label{sec:spbg}

\subsection{Review of spontaneous baryogenesis}

First, we shall explain how spontaneous baryogenesis works.  For
simplicity, we assume that the system has only baryon symmetry,
$U(1)_B$.  The extension to the case with several global $U(1)$
symmetries will be discussed later.  Let us consider a scalar field
$a$, which is derivatively coupled to the baryon current:
\begin{equation}
\label{eq:der_int}
{\cal L}_{eff} = - \frac{\del_\mu a}{M} J_B^\mu,
\end{equation}
where $M$ is a cutoff scale. The baryonic current is given by
\bea
J_B^\mu &=&
\sum_i B_i j^\mu_i,\non\\
j_i^\mu &=& \left\{
\bear{ll}
 \overline{\psi_i} \gamma^\mu \psi_i  & {\rm for~~fermions},\\
&\\
i \left(\varphi_i
 \del^\mu \varphi_i^*-\varphi_i^* \del^\mu \varphi_i
 \right) &
 {\rm for~~bosons},
\eear
\right.
\eea
where $B_i$ is the baryon number of the $i$th field. If $\del_\mu a$
takes a nonvanishing classical value, the above interaction induces
spontaneous CPT violation since $\del_\mu a$ is odd under CPT
transformations. Assuming that the scalar field $a$ is homogeneous, we
have
\begin{eqnarray}
{\cal L}_{eff} &=& - \frac{\dot{a}}{M} n_B \non\\
                      &\equiv& \sum_i \mu_i n_i,
\end{eqnarray}
where the overdot denotes differentiation with respect to time, and $n_B$
is the baryon number density.  Also we define the chemical potential
$\mu_i \equiv - \dot{a} B_i/M$, and the number density of the $i$th
field, $n_i\equiv j_i^0$. Thus, if $U(1)_B$ violating operators are in
thermal equilibrium, as the thermal and chemical equilibrium states
change, the number density $n_i$ for the $i$th field follows the
chemical potential $\mu_i$.

When the baryon number violating interactions decouple at last, the
baryon number density is frozen at the value at that time \cite{CK}:
\beq
\label{eq:nd}
n_{B}(t_D) = \sum_i B_i\frac{g_i  \kappa_i  T_D^3}{6} \left\{
\frac{\mu_i}{T_D} + O\left[\left(\frac{\mu_i}{T_D}\right)^3\right]
\right\}\,,
\eeq
where $T_D$ is the decoupling temperature and $g_i$ represents the
degrees of freedom of the corresponding field.  Also, $\kappa_i$ is
defined as
\beq
\kappa_i =\left\{
\begin{array}{ll}
1&{\rm for~~fermions}, \\
2&{\rm for~~bosons}.
\end{array}
\right.
\eeq
It should be emphasized that such asymmetries are realized only in the
light fields that contribute to the energy of the universe as
radiation.

\subsection{Nambu-Goldstone bosons associated with flat directions}

Here we show that  derivative interactions like
Eq.~(\ref{eq:der_int}) are naturally present in the minimal
supersymmetric standard model, which contains many flat
directions.  Since flat directions are composed of the standard model
fields charged under several (global or local) $U(1)$ symmetries, the
NG bosons associated with these symmetries are induced if a flat
direction develops a nonzero vacuum expectation value (VEV). It is
one of these NG bosons that realizes spontaneous baryogenesis through
derivative interactions. The purpose of this section is to identify
this NG boson, and derive the derivative interaction relevant for 
spontaneous baryogenesis.

A flat direction $X$ is specified by a holomorphic gauge-invariant
polynomial:
\beq
\label{eq:monomial}
X \equiv \prod_{i}^{N} \chi_i,
\eeq
where $N$ superfields $\left\{ \chi_i \right\}$ constitute the flat
direction $X$, and we have suppressed the gauge and family indices
with the understanding that the latin letter $i$ contains all the
information to label those constituents.  When $X$ has a nonzero
expectation value, each constituent field also takes a nonzero
expectation value
\begin{equation}
\label{eq:chivev}
\la \chi_i \ra = \frac{f_i}{\sqrt{2}} e^{i \theta_i}
               = \frac{f_i}{\sqrt{2}} e^{i P_i/f_i}.
\end{equation}
Here each $f_i/\sqrt{2}$ is the absolute value of the expectation
value $\la \chi_i \ra$, and is related to every other due to the
D-(F-)flat conditions. $P_i$ is a canonically normalized field
corresponding to the phase component $\theta_i$.  For the moment, we
will pay attention to the baryon $(B)$ and lepton $(L)$ symmetries,
and consider flat directions whose expectation values break both
symmetries.\footnote{ Strictly speaking, we must take into account the
  weak hypercharge $Y$.  However, it does not change our result, so we
  neglect its effect for simplicity.  }. Later we consider the case
that the standard model particles are charged under extra global
$U(1)$ symmetry, such as the $PQ$ symmetry.  We define $U(1)_{A \pm}$,
which are the two independent linear combinations of $U(1)_{B}$ and
$U(1)_{L}$ symmetries:
\begin{eqnarray}
\label{eq:def+-}
U(1)_{A +} &:& Q^+ =  \cos{\xi} ~B + \sin{\xi} ~L,\non\\
U(1)_{A -} &:& Q^- =  -\sin{\xi} ~B + \cos{\xi} ~L,\non\\
 \tan{\xi} &\equiv& \frac{\sum_i L_i}{\sum_i B_i}.
\end{eqnarray}
Here $B$, $L$, and $Q^\pm$ are the charges of $U(1)_{B}$, $U(1)_{L}$,
and $U(1)_{A \pm}$, respectively.  We define the generators of
$U(1)_{B}$, $U(1)_{L}$, and $U(1)_{A \pm}$ as $\Upsilon_{B}$,
$\Upsilon_{L}$, and $\Upsilon_{A \pm}$.
Following the argument in Refs.~\cite{sikivie,DF},  we express $\theta_i$ as
\bea
\theta_i &\equiv&  B_i \alpha_B + L_i \alpha_L \non\\
&\equiv& Q_i^+ \alpha_+ + Q_i^- \alpha_-,
\eea
where $B_i$, $L_i$, and $Q_i^\pm$ are the corresponding charges of
$\chi_i$, and $\alpha_{B}$, $\alpha_{L}$, and $\alpha_{\pm}$ are the
angles conjugate to the generators $\Upsilon_{B}$, $\Upsilon_{L}$, and
$\Upsilon_{A \pm}$, respectively.  They are related as follows:
\beq
\left(\bear{c}
\alpha_B\\
\alpha_L
\eear
\right) =\left(
\bear{cc}
\cos{\xi}&-\sin{\xi}\\
\sin{\xi}&\cos{\xi}
\eear
\right)
\left(
\bear{c}
\alpha_+\\
\alpha_-
\eear
\right).
\eeq
The NG bosons associated with the spontaneous breaking of 
$U(1)_{B}$, $U(1)_{L}$, and $U(1)_{A \pm}$ are denoted as $a_B$, $a_L$,
and $a_\pm$.  Finally, we define the $2 \times 2$ real matrix of decay
constants $F$ as
\beq
\left(\bear{c}
a_+\\
a_-
\eear
\right) =F
\left(
\bear{c}
\alpha_+\\
\alpha_-
\eear
\right).
\eeq
Since the kinetic terms of $a_\pm$ come from those of  ${\chi_i}$
with ${f_i}$ fixed, $F$ is related to the amplitudes ${f_i}$:
\beq
\label{eq:ftf}
F^T F = \sum_i f_i^2
\left(
\bear{cc}
 Q_i^+ Q_i^+&Q_i^+ Q_i^-\\
 Q_i^- Q_i^+&Q_i^- Q_i^-\
\eear
\right).\eeq
As shown below, $a_+$ becomes massive due to the existence of the
A-term, while $a_-$ remains massless.  In order to solve for $F$, it
is necessary to know the effective potential for flat directions.

Flat directions are lifted by both supersymmetry-breaking effects and
nonrenormalizable operators, although there are no classical
potentials along flat directions in the supersymmetric limit. Since a
flat direction is well described by a single complex scalar field
$\Phi \equiv \phi/\sqrt{2}\, e^{i \theta}$, these effects induce a
potential for $\Phi$.  Here $\phi$ is the amplitude, whose expectation
value is given by a typical value of $f_i$, and the phase $\theta$ is
defined as
\begin{equation}
\label{eq:def_of_theta}
\theta \equiv \frac{1}{N} \sum_i  \theta_i 
= \frac{1}{N} \sum_i  Q_i^+ \alpha_+.
\end{equation}
So the scalar field $\Phi$ is often defined as $\Phi^N \equiv X$ (as
shorthand).  Assuming gravity-mediated supersymmetry breaking, the flat
direction is lifted by supersymmetry-breaking
effects~\cite{EnqvistMcDonald98},
\begin{equation}
 \label{eq:grav}
    V_{\rm grav} \simeq m_{\phi}^2 \left[ 1+K
      \log \left(\frac{|\Phi|^2}{M_G^2} \right)\right] |\Phi|^2,
\end{equation}
where $m_\phi \sim 1$TeV is a soft mass, $K$ is a numerical
coefficient of one-loop corrections, and $M_G$ is the reduced Planck
mass. Moreover, assuming a nonrenormalizable operator in the
superpotential of the form
\begin{equation}
\label{eq:spnr}
    W_{\rm NR} = \frac{X^k}{N k M^{N k-3}} =  \frac{\Phi^n}{n M^{n-3}} \,,
\end{equation}
the flat direction is further lifted by the potential
\begin{equation}
\label{eq:pnr}
    V_{\rm NR} = \frac{|\Phi|^{2 n-2}}{M^{2 n -6}}\,,
\end{equation}
where $n \equiv N k$ and $M$ is a cutoff scale. In fact, the
nonrenormalizable superpotential not only lifts the potential but
also gives the baryon and/or lepton number violating A-terms of the
form
\begin{eqnarray}
\label{eq:nrA}
    V_{As} &=& a_m \frac{m_{3/2}}{N k M^{N k -3}} X^k + {\rm h.c.}\,\non\\
    &=& a_m \frac{m_{3/2}}{n M^{n -3}} \Phi^n + {\rm h.c.}\,,
\end{eqnarray}
where $m_{3/2}$ is the gravitino mass, $a_m$ is a complex constant of
order unity, and we assume a vanishing cosmological constant. With the
redefinition of the phase of $\Phi$, $a_m$ can be real. Hereafter we
adopt this convention.

On the other hand, it is also possible that nonrenormalizable
operators in Eq.~(\ref{eq:spnr}) are forbidden by some
R-symmetries. Then the effective potential is parabolic like
Eq.~(\ref{eq:grav}) up to some cutoff scale of the K\"ahler potential,
$M_*$. The A-terms are also supplied by the K\"ahler potential as
\begin{eqnarray}
\label{eq:KA}
    V_{Ak} &=& a_* \frac{m_{3/2}^2}{N k M_*^{N k -2}} X^k + {\rm h.c.}\,\non\\
    &=&  a_* \frac{m_{3/2}^2}{n M_*^{n -2}} \Phi^n + {\rm h.c.}\,,
\end{eqnarray}
where $a_*$ is taken to be a real constant of order unity.  Note that
the dependence of the A-term on the gravitino mass is different from
that of the nonrenormalizable term in the superpotential.

During the inflationary epoch, a flat direction can easily acquire a
large expectation value. Strictly speaking, if a flat direction has a
four-point coupling to the inflaton in the K\"ahler potential with
appropriate sign and magnitude, it has a negative mass squared
proportional to the Hubble parameter squared \cite{DRT},
\begin{equation}
\label{eq:hmass}
   V_H = -c_H H^2 |\Phi|^2\,,
\end{equation}
where $c_H$ is a positive constant of order unity. This negative mass
term destabilizes the flat direction at the origin, and the flat
direction rolls down toward the minimum of the potential.  The
position of the minimum depends on whether the nonrenormalizable
superpotential exists or not. If it exists, the minimum is determined
by the balance between the negative mass term $V_H$ and the
nonrenormalizable potential $V_{\rm NR}$. If not, it is fixed around
the cutoff scale $\sim M_*$.  Thus, the minimum of the potential
during inflation is given by
\beq
\label{eq:potential_min}
\phi_{\rm min} \sim \left\{
\begin{array}{ll}
\left(H M^{n-3} \right) ^{\frac{1}{n-2}}
& {\rm with~~}W_{\rm NR}, \\
M_* 
& {\rm without~~}W_{\rm NR}.
\end{array}
\right.
\eeq
Notice that the flat direction takes the nonzero expectation value
given by Eq.~(\ref{eq:potential_min}) even after inflation ends, once
the initial position is set during inflation. We would like to comment
on thermal effects on flat directions. Since those fields that couple
to the flat direction obtain a large mass of $O(f \phi_{\rm min})$ where
$f$ represents the yukawa or gauge coupling constant, they must be
out-of-equilibrium. Therefore we do not take account of thermal effects
in the following.

Now we consider the implication of the A-term.  As one can see, the
phase $\theta_i$ appears only in the A-term. In either case, it can be
expressed as
\beq
\label{eq:genaterm}
V_A = M_A^4 \cos{\left[
k {\cal Q} \alpha_+
\right]},
\eeq
where 
\beq
{\cal Q} \equiv \sqrt{\left(\sum_i B_i \right)^2+\left(\sum_i L_i \right)^2}
\eeq
and $M_A$ denotes the energy scale of the A-term.  This means that
$U(1)_{A+}$ is explicitly violated by the A-term, while $U(1)_{A-}$
remains intact. Since $a_+$ becomes massive due to the interaction in
Eq.~(\ref{eq:genaterm}), $F$ takes the following form:
\beq
\label{eq:def_of_F}
F = \left(
\bear{cc}
v_a&0\\
f_{01}&f_{11}
\eear
\right).
\eeq
Substituting this equation into Eq.~(\ref{eq:ftf}), we have
\bea
v_a^2 &=& \sum_i f_i^2 Q_i^+{}^2 - \frac{\left( \sum_i f_i^2 Q_i^+ Q_i^-\right)^2}{
 \sum_i f_i^2 Q_i^-{}^2} ,\non\\
 f_{01} &=& \frac{\sum_i f_i^2 Q_i^+ Q_i^-}{
\sqrt{ \sum_i f_i^2 Q_i^-{}^2}},\non\\
f_{11} &=& \sqrt{\sum_i f_i^2 Q_i^-{}^2},
\eea
where we have assumed that the $Q_i^{-}$s are not all zero.
Thus we can identify the NG boson that becomes massive due to the
existence of the A-term:
\bea
a_+ &=& v_a \alpha_+ \non\\
       &=& v_a \frac{\theta}{\frac{1}{N} \sum_i Q_i^+}, 
\eea
where we used Eq.~(\ref{eq:def_of_theta}) in the last equality.  This
equation illuminates the reason why a flat direction can be well
described by a single complex scalar field $\Phi$. Its amplitude
represents the magnitude of the expectation value of the flat
direction, and the phase corresponds to the dynamical NG mode whose
motion is affected by the A-term.  The equation of motion of $a_+$ is
given by
\beq
\label{eq:eom}
\ddot{a}_+ + 3 H \dot{a}_+-
M_A^4 \frac{k {\cal Q}}{v_a} \sin{\left(k {\cal Q} \frac{a_+}{v_a}\right)} =0,
\eeq
where we assumed that the $\{f_i \}$ are constant.\footnote{As long as
  $\dot{f_i} \lesssim H f_i$, the equation of motion is still valid up
  to a numerical factor in the viscosity term.} Its velocity can be
  estimated as
\beq
\label{eq:vel}
\left|\dot{a}_+ \right| \sim \frac{k {\cal Q}}{H v_a} M_A^4,
\eeq
where we used the slow-roll approximation because the inverse
curvature scale of the potential is roughly $\sqrt{m_{3/2} H} \ll
H$.\footnote{
During inflation, the effective mass of $a_+$ is
comparable to the Hubble scale due to the Hubble induced A-term so
that quantum fluctuations of $a_+$ are negligible.
 }  We have also
assumed that $a_+$ remains far from the extremum of the potential by
$O(v_a)$.  The velocity plays an essential role for both the AD
mechanism and spontaneous baryogenesis in our scenario.  In fact, it
generates the asymmetry of  $U(1)_{A +}$ as a condensate of the
flat direction.  The $Q^+$ number density $n_+$ can be calculated as
\bea
\label{eq:asym}
n_+  &=&  -\sum_i Q_i^+ f_i^2 \dot{\theta_i} \non\\
         &=& -v_a \dot{a}_+,
\eea
while the $Q^-$ number density $n_-$ remains zero.  In the AD
mechanism, the baryon (or lepton) asymmetry comes from this $U(1)_{A
  +}$ asymmetry.

With this background we turn now to an account of the derivative
interactions.  The NG mode $a_+$ transforms as $a_+ \rightarrow a_+ +
v_{a} \epsilon$ under the $U(1)_{A+}$ transformation $\alpha_+
\rightarrow \alpha_+ + \epsilon$.  Since the $\left\{\chi_i \right\}$
are the particles of the standard model, they participate in the
following Yukawa superpotential of the standard model:
\begin{eqnarray}
\label{eq:smsuper}
W_{\rm SM} &=&y_u u\, Q \,H_u + y_d d\, Q \,H_d
             +y_e e\, L \,H_d,
\end{eqnarray}
where $Q$ and $L$ are $SU(2)_L$ doublet quarks and leptons, $u$, $d$, and
$e$ are $SU(2)_L$ singlet quarks and leptons, and $H_u$($H_d)$ is the
up-(down-)type Higgs superfield.  The Yukawa interactions above therefore
depend on the NG mode $a_+$ when the flat direction has an expectation
value. In order to obtain the interaction between the NG mode $a_+$ and
the other charged fields, we define the $U(1)_{A+}$ current as
\beq
  J^{\mu}_{A+} \equiv -\sum_{m'} \frac{\del\CL}{\del(\del_{\mu} \chi_{m'})} 
                      \delta\chi_{m'},
\eeq
where $m'$ denotes all fields with nonzero $U(1)_{A+}$ charges, that is,
$\chi_{m'}$ transforms under $U(1)_{A+}$ symmetry as $\chi_{m'}
\rightarrow \chi_{m'} + \epsilon \delta\chi_{m'}$ with $\delta\chi_{m'} = i
Q^{+}_{m'} \chi_{m'}$. Then, the $U(1)_{A+}$ current is given by
\beq
  J^{\mu}_{A+} = v_{a} \del^{\mu} a_+ + \sum_{m} Q^{+}_{m} j^\mu_m,
\eeq
where $m$ denotes all fields with nonzero $U(1)_{A+}$ charges except
the NG mode $a_+$. Current conservation yields the equation of
motion for $a_+$
\beq
  \del_{\mu} J^{\mu}_{A+} = v_{a} \del^2 a_{+} 
                      + \sum_{m} Q^{+}_{m} \del_{\mu} j^\mu_m = 0.
\eeq
Here the first term in the middle equation is the kinetic term for the
NG mode $a_+$ and the second term can be derived from the following
effective Lagrangian,
\beq
  \CL_{\rm eff} = -\sum_{m} \frac{Q^{+}_{m}}{v_a}
                   \left( \del_\mu a_+ \right) j^\mu_m,
  \label{eq:dercoupling}
\eeq
which yields the derivative interactions between the NG mode $a_+$ and
the other charged fields. Although the index $m$ should be taken over
all the charged fields except the NG mode $a_+$, we will concentrate
on a derivative coupling of all light fields because the asymmetries
are induced only in the light degrees of freedom.  Hereafter the
subscript $m$ denote the species of light fields unless
otherwise stated. Note that the above discussion applies for the
$U(1)_{A-}$ symmetry in the same way, and that there also exists the
derivative coupling of $a_-$ with the $U(1)_{A-}$ current. However, it
does not have any significant meaning, since $\dot{a}_-$ cannot
have a nonzero classical value.

Up to this point we have considered the case that the system has only
baryon and lepton symmetries. However, the essential points of our
argument expressed so far are still valid in the general case with
extra $U(1)$ symmetries. There exists only one NG boson which becomes
massive due to the A-term. In order to extract it, we may need to take
a superposition of the symmetries as shown in Eq.~(\ref{eq:def+-}).
It is always possible to rotate out the NG boson from the Yukawa
interactions, leading to a derivative interaction with a current. In
particular, Eqs.~(\ref{eq:asym}) and (\ref{eq:dercoupling}) are valid
with the understanding that $U(1)_{A+}$ is the symmetry violated by
the A-term, and that $Q^+$ and $a_+$ represent the corresponding
charge and NG boson, respectively.

\subsection{Symmetry breaking operators in thermal equilibrium}

We would next like to focus on the subtleties associated with the fact
that there are several $U(1)$ symmetries. In the simple case
illustrated in the first part of this section, thermal and chemical
equilibrium are attained for each field, with the chemical potential
given by the coefficient of the number density in the derivative
interaction. In general, this is not the case. The asymmetries
generated through spontaneous baryogenesis crucially depend on the
symmetry-breaking operators in thermal equilibrium.  For simplicity,
we consider the case with baryon and lepton symmetries.
Assuming that sphaleron processes are effective at a later epoch, we
concentrate on $B-L$ violating operators in thermal equilibrium. The
final $B-L$ asymmetry is determined by the one with the lowest
decoupling temperature,\footnote{ Here we assume that the decoupling
temperatures are not degenerate.} as long as $\mu T^2$ decreases more slowly
than $a^{-3}$.  Such an interaction can be characterized by the amount
of $B$ and $L$ violation, $\Delta_B$ and $\Delta_L$, respectively. Since
the baryon and lepton asymmetries are generated through this
interaction,\footnote{ Hereafter we assume that the interactions which
transmit the generated baryon and lepton numbers to other particles are
in thermal equilibrium.} they are related as $n_B \Delta_L = n_L
\Delta_B$, that is,
\bea
\label{eq:constraint}
 \sum_m \Xi_m n_m &=&0,
\eea
where we defined $\Xi_m = B_m \Delta_L-L_m \Delta_B$.  We would like
to know the resultant $n_m$ induced by the derivative interaction
Eq.~(\ref{eq:dercoupling}) under this constraint.  If it were not for
the constraint, the following asymmetry would be generated:
\beq
\overline{n}_m (t_D) = \frac{ \kappa_m g_m}{6} \overline{\mu}_m T_D^2,
\eeq
where $\overline{\mu}_m \equiv - Q^+_m \dot{a}_+/v_a$.  It is worth
noting that $\overline{\mu}_m$ cannot be interpreted to be the
chemical potential of the $m$th field, if we take the constraint into
account. The reason for this is not difficult to see. Due to the
constraint, $n_m$ is not able to vary freely. In fact, only the
projection of $\{\overline{\mu}_m\}$ onto the parameter plane
perpendicular to $\{\Xi_m\}$ has physical meaning.  That is to say,
the resultant asymmetry should depend on $\tilde{\mu}_m$ defined as
\beq
\label{eq:mutilde}
\tilde{\mu}_m
\equiv \overline{\mu}_m -
\frac{\left(\overline{\mu} \cdot \Xi \right)}{\Xi^2}
\Xi_m,
\eeq
where we adopt the following shorthand. 
\bea
Y^2 &\equiv& \sum_m \kappa_m g_m Y_m^2,\non\\
Y \cdot Z &\equiv& \sum_m \kappa_m g_mY_m Z_m.
\eea
Then it is easy to show that $\tilde{\mu}_m$ are invariant under
the transformation $\overline{\mu}_m \rightarrow \overline{\mu}_m +
\alpha \Xi_m$ for an arbitrary constant $\alpha$. If we require that
$\{n_m\}$ take the form of thermal and chemical equilibrium, the
$\{n_m\}$ are then uniquely determined by using $\tilde{\mu}_m$:
\beq
n_m (t_D) =
 \frac{ \kappa_m g_m}{6} \tilde{\mu}_m T_D^2.
\eeq
The resultant $B-L$ number density is then given by
\bea
\label{eq:b-l}
n_{B-L}(t_D) &=& \sum_m (B_m-L_m) n_m \non\\
                     &=& 
                     \left(\Delta_B - \Delta_L \right)
                     \left(\mu_{B} \Delta_B + \mu_{L} \Delta_L \right)\non\\
                     && \times
                  \frac{  
                        B^2 L^2
        }{
                        B^2 \Delta_L^2 
                   + L^2 \Delta_B^2                 
                     } 
                     \frac{T_D^2}{6},
\eea
where $\mu_B \equiv - \cos{\xi}~\dot{a}_+/v_a$ and $\mu_L \equiv
-\sin{\xi}~\dot{a}_+/v_a$, so that $\overline{\mu}_m = \mu_B B_m +
\mu_L L_m$. Thus, the following two conditions must be met for 
$B-L$ asymmetry to be generated:
\bea
\Delta_B - \Delta_L &\ne& 0, \\
\mu_{B} \Delta_B + \mu_{L} \Delta_L &\ne& 0.
\eea
The meaning of the first condition is clear. The interaction in
thermal equilibrium should, of course, violate $B-L$ symmetry. In
order to understand the second condition, we rewrite it as follows:
\bea
  \mu_{B} \Delta_B + \mu_{L} \Delta_L &=& 
   - (\Delta_B \cos{\xi} + \Delta_L \sin{\xi}) \frac{\dot{a}_+}{v_a} 
       \non  \\
    &=&- \Delta_{Q_{+}} \frac{\dot{a}_+}{v_a}.
\eea
Hence, the second one means that  interaction in thermal equilibrium
must also violate  $U(1)_{+}$ symmetry, so that
the derivative interaction does induce some asymmetries.  If
$\Delta_{Q_+} = 0$, no asymmetries result since the broken symmetry in
thermal equilibrium is then orthogonal to the $U(1)_{+}$ symmetry, for
which nonzero chemical potential is induced by the derivative
interaction. What needs to be emphasized at this juncture is that the
asymmetries are induced even for  fields that do not participate
in derivative interactions. The reason for this is that the
constraint Eq.~(\ref{eq:constraint}) relates the fields with
$\overline{\mu}_m \ne 0$ to those with $ \overline{\mu}_m = 0$, so that
the latter can feel the effective chemical potential. For example, it
is even possible to set all the derivative interactions in the hidden
sector, as long as there exists an interaction in thermal equilibrium
that violates both the hidden symmetry and $B-L$ symmetry. Then the
asymmetry generated in the hidden sector also induces  $B-L$
asymmetry through such an interaction. We shall detail an application of
this striking feature in the next section.

\section{Applications}
\label{sec:applications}
\subsection{Baryogenesis in flat directions with $B-L=0$}
As an application of the spontaneous baryogenesis discussed in the
previous section, here we consider baryogenesis in those flat
directions with $B-L=0$ and $B+L \ne 0$. In the next subsection we
consider  baryogenesis in flat directions with $B=L=0$.  The AD
mechanism generates the asymmetry of the $U(1)$ charge, which is
explicitly violated by the A-term. Hence it does not work for flat
directions with $B-L=0$, unless $Q$-balls are formed.  Even if $Q$-balls
are formed, they might decay and/or evaporate away before the
electroweak phase transition. Then no baryon asymmetry results due to
the sphaleron processes.  As listed in Ref. \cite{tony}, in fact, there
are many flat directions that have vanishing $B-L$ (e.g., $QQQL$,
$uude$), and some of them have neither baryon nor lepton charges (for
example, $QuQd$, $QuLe$).  If such flat directions were selected during
inflation, it is difficult to explain the baryon asymmetry in the
present universe. Thus, it is intriguing to examine whether it is
possible to realize baryogenesis even when a flat direction with $B-L=0$
is selected.

Since we are considering the flat direction with
$B=L\ne0$, $Q^{\pm}$ and $v_a$ defined in Eqs.~(\ref{eq:def+-}) and
(\ref{eq:def_of_F}) are given by
\bea
Q^+ &=& \frac{B+L}{\sqrt{2}},\non\\
Q^- &=& \frac{-B+L}{\sqrt{2}},\non\\
v_a^2 &=& 2 \frac{\sum_{i,j}f_i^2 f_j^2 B_i^2 L_j^2}{\sum_i f_i^2 (B_i^2 + L_i^2)}\sim\phi_{\rm min}^2,
\eea
where we have used $f_i \sim \phi_{\rm min}$ and $B_i \sim L_i \sim 1$
in the last line.  The derivative interaction leads to the following:
\bea
{\cal L}_{\rm eff} &=& \sum_i \overline{\mu}_i n_i,
\eea
with
\beq
\overline{\mu}_i = -
   \frac{B_i+L_i}{\sqrt{2} v_a} \dot{a}_+ 
\sim - \frac{k {\cal Q}}{\sqrt{2}} m_{3/2} (B_i+L_i),  
\eeq
where we have used Eqs.~(\ref{eq:potential_min}) and (\ref{eq:vel})
and assumed the existence of a  nonrenormalizable term and that
$\dot{a}_+ > 0$.

We need to incorporate a $B-L$ breaking operator in thermal
equilibrium that also violates $B+L$ symmetry, in order to realize 
spontaneous baryogenesis. The resultant $B-L$ asymmetry can be
estimated as Eq.~(\ref{eq:b-l}) with
\beq
\mu_B = \mu_L = - \frac{k {\cal Q}}{\sqrt{2}} m_{3/2}.
\eeq
As an illustration, we consider
the following $B-L$ breaking operator:
\begin{equation}
\label{eq:dim5}
{\cal L}_{\not{L}} = \frac{2}{v}l\, l\, H_u H_u +\, {\rm h.c.},
\end{equation}
where $v$ is the scale characterizing the interaction and may be 
identified with the heavy Majorana mass for the right-handed neutrino
in the context of the seesaw mechanism. This interaction violates the
lepton number by two, while the baryon number is intact: $\Delta_L = 2$ and
$\Delta_B =0$, so we have $\Xi_m = 2 B_m$.  
Then the effective chemical
potential $\tilde{\mu}_m$ can be calculated as
\bea
\tilde{\mu}_m 
&\simeq& - \frac{k {\cal Q}}{\sqrt{2}} m_{3/2} L_m.
\eea
Naturally, this result means that only lepton asymmetry is
generated through  spontaneous baryogenesis.  Thus the $B-L$
asymmetry at decoupling is
\beq
\frac{n_{B-L}}{s}=\frac{15 k {\cal Q}}{4 \sqrt{2}\pi^2} 
\frac{L^2
m_{3/2}}{g_* T_D},
\eeq
where $g_*$ counts the effective degrees of freedom for relativistic
particles.  The baryon asymmetry is obtained through the sphaleron
effects\footnote{
If the weak gauge bosons become massive due to the large VEV of the
flat direction, the sphaleron configuration is not excited before the decay of
the flat direction. Even if the flat direction is composed of $SU(2)$ singlet fields,
the sphaleron process might be out of equilibrium at decoupling, since
the decoupling temperature is rather high ($\sim 10^{12}$ GeV as shown below).
Hereafter we assume that this is the case.  
}
as~\cite{KSHT}
\beq
\frac{n_{B}}{s}=\frac{30 k {\cal Q}}{23 \sqrt{2}\pi^2}
 \frac{L^2
m_{3/2}}{g_* T_D}.
\eeq
In order to estimate the baryon asymmetry, we need to know the typical
value of $T_D$.

The lepton number violating  rate of the interaction given by Eq.~(\ref{eq:dim5}) is
given by $\Gamma \sim 0.04 T^3 / v^2$ \cite{sarkar}.
Then, the decoupling temperature is calculated as
\begin{eqnarray}
\label{eq:decT}
T_D &\sim&5 \times 
10^{11} {\rm GeV} \left(\frac{g_*}{200}\right)^{\frac{1}{2}}
                    \left(\frac{v}{10^{14} {\rm GeV}}\right)^2,\non\\
   &\sim & 7 \times 10^{11} {\rm GeV} 
                     \left(\frac{g_*}{200}\right)^{\frac{1}{2}}
                     \left(\frac{m_\nu}{1{~\rm eV}}\right)^{-2}
                     \sin^4 \beta, \non\\
\end{eqnarray}
where $\tan \beta \equiv \la H_u \ra/\la H_d \ra$, and $m_\nu$
is the neutrino mass related to $v$ by $m_\nu = 4 \la H_u \ra^2 /v$.
Also we have assumed that the reheating process ends before 
the decoupling. 
Such a high reheating temperature might lead to the gravitino
problem~\cite{gra}. For $m_{3/2} = 3 \sim 10$ TeV, the reheating
temperature is constrained as $T_{RH} \lesssim 10^{12}$ GeV, assuming
that the mass of the lightest supersymmetric particle (LSP) is $O(100$
GeV) \cite{gra}. If we take this bound seriously, the decoupling
temperature $T_D$ must be less than $10^{12}$ GeV so that the lepton
number violating interaction is in thermal equilibrium. For $m_{3/2}
\sim 100$ GeV, the reheating temperature must be smaller than $T_{RH}
\lesssim 10^{9}$ GeV.
Alternatively, these constraints on the reheating temperature can be
evaded by the introduction of a supersymmetric partner with a mass
much lighter than $100$ GeV. One such particle is the axino.  In fact,
it was shown that the reheating temperature is constrained rather
loosely as $T_{RH} < 10^{15}$ GeV for $m_{3/2} \simeq 100$ GeV, if the
axino is the LSP and the gravitino is the next LSP \cite{asaka}.

As long as the gravitino problem is absent, the typical value of the
decoupling temperature is $T_D \sim 10^{12}$ GeV.  Hence we obtain the
right amount of baryon asymmetry,
\bea
\frac{n_{B}}{s} &\sim& 0.1 \frac{m_{3/2}}{T_D},\non\\
&\simeq& 3\times10^{-10} \left(\frac{m_{3/2}}{3 {\rm TeV}}\right) 
\left(\frac{T_D}{10^{12} {\rm GeV}}\right)^{-1}\,.
\eea

\subsection{Baryogenesis in flat directions with $B=L=0$}

Next we show that baryogenesis is possible even for flat directions
with $B=L=0$. Our strategy is as follows. Since $\sum_i B_i =\sum_i
L_i =0$, the baryon and lepton symmetries are not violated by the
A-term.  Therefore the corresponding NG bosons $a_B$ and $a_L$ remain
massless. Note that there is a degree of freedom $a$ which becomes
massive due to the A-term and develops a classical value $\dot{a} \ne
0$.  However, it is not a NG boson since the system does not possess
the corresponding symmetry, so it cannot have any derivative
interactions with the currents. Now it is clear what must be added to
the system. We need to incorporate an additional $U(1)$ symmetry which
is explicitly violated by the A-term. Then the corresponding NG boson
obtains a nonzero velocity, leading to the derivative interactions
relevant for  spontaneous baryogenesis.

In order to realize our scenario, the standard particles should be charged
under another global symmetry in addition to the baryon and lepton
symmetries. One of the famous examples is $PQ$ symmetry, which was
introduced to solve the strong $CP$ problem of quantum chromodynamics
\cite{pq}. For definiteness, we adopt the supersymmetric Dine-Fischler-Srednicki-Zhitnitsky (DFSZ) axion
model\footnote{
  Here we assume that the $PQ$ scalar fields, which are responsible
  for the spontaneous $PQ$ symmetry breaking in the present universe,
  have negative mass squared of order $H$ during inflation. Then we
  can avoid the problem of axion domain walls. In addition, cold dark matter can
  also be explained by the axion in our scenario as a by-product of
  adopting the $PQ$ symmetry.
} \cite{DFSZ,Ma}, but it is
trivial to extend it to the case with general global $U(1)$
symmetries. Then, a flat direction with neither baryon nor lepton charge
can have nonzero $PQ$ charges if we assign them properly to standard particles.
 In the same way as before, we express $\theta_i$ as
\beq
\theta_i = R_i \alpha_R + B_i \alpha_B + L_i \alpha_L,
\eeq
where $R_i$ is the $PQ$ charge of $\chi_i$, and $\alpha_R$ is the
angle conjugate to the generator $\Upsilon_R$. Then the A-term 
can be written as
\beq
V_A = M_A^4 \cos\left[
k {\cal R} \alpha_R
\right],
\eeq
where ${\cal R} \equiv \sum_i R_i$. We define the NG boson corresponding
to  $PQ$ symmetry as
\beq
a_R = v_a \alpha_R,
\eeq
where $v_a$ now takes the complicated form~\cite{sikivie}
\bea
\label{eq:var}
v_a^2 &=& 
\sum_i R_i^2 f_i^2 - \frac{
\left(\sum_{i} R_i B_i f_i^2\right)^2
}{\sum_i B_i^2 f_i^2}\non\\
&&~~~~~~~~~~~~~- \frac{
\left(\sum_{i} R_i L_i f_i^2\right)^2
}{\sum_i L_i^2 f_i^2}\non\\
&\sim& \phi_{\rm min}^2.
\eea
We have assumed that $\phi_{\rm min}$ is much larger than the breaking 
scale of the $PQ$ symmetry, $F_a$, in the present universe. In fact, the NG boson
$a_R$ continuously transforms into the axion that is the phase component
of the $PQ$ scalar field, after the flat direction starts to oscillate.
Differentiation of $a_R$ with respect to time is likewise estimated as
\beq
\label{eq:adot}
\left| \dot{a}_R \right| \sim \frac{k {\cal R}}{H v_a} M_A^4.
\eeq
The time component of the derivative interaction then reads
\beq
{\cal L}_{\rm eff}=\sum_m \overline{\mu}_m n_m
\eeq
with
\beq
\overline{\mu}_m = -\frac{R_m}{v_a} \dot{a}_R 
           \sim - k {\cal R} R_m m_{3/2},
\eeq
where we have used Eqs.~(\ref{eq:var}) and (\ref{eq:adot}), and assumed
the existence of a nonrenormalizable superpotential and $\dot{a}_R > 0$.

In order to estimate the resultant number density, we must take into
account two constraints like Eq.~(\ref{eq:constraint}), because the
system now possesses three $U(1)$ symmetries. We assume that the $B-L$
violating interaction\footnote{
  The $PQ$ symmetry is also violated by strong sphaleron
  processes.  However, the strong sphaleron configuration is not
  excited as long as the $SU(3)_C$ is spontaneously broken by the
  large VEV of the flat direction.  Fortunately, since all the $B=L=0$
  directions contain squarks, this is always the case.  Furthermore,
  the $\mu$-term, which breaks $PQ$ symmetry, is out of equilibrium at
  $T_D \sim 10^{12}$ GeV. Hereafter we assume that the other
  interactions violate neither $PQ$ nor $B-L$ symmetry at decoupling.
} in thermal equilibrium breaks the $PQ$, $B$, and $L$ symmetries by $\Delta_R$,
$\Delta_B$, and $\Delta_L$, respectively.
The constraints are written as
\beq
\sum_m \Xi_m^{(i)} n_m = 0~~~~{\rm for~~}i=1,~2,~3,
\eeq
\bea
\Xi_m^{(1)} &\equiv&R_m \Delta_L-L_m \Delta_R,\non\\
\Xi_m^{(2)} &\equiv&R_m \Delta_B-B_m \Delta_R,\non\\
\Xi_m^{(3)} &\equiv&B_m \Delta_L-L_m \Delta_B,\non\\
\eea
and two of them are independent. The resultant number density is given by
\beq
n_m(t_D) = \frac{\kappa_m g_m}{6} \tilde{\mu}_m T_D^2,
\eeq
where
\bea
\tilde{\mu}_m &\equiv& \overline{\mu}_m 
                       - \beta^{(1)}\Xi^{(1)} - \beta^{(2)}\Xi^{(2)} ,\\
\beta^{(1)} &\equiv& 
         \frac{ \Xi^{(2)}{}^2 
                \left( \overline{\mu} \cdot \Xi^{(1)} \right)
               - \left( \Xi^{(1)} \cdot \Xi^{(2)} \right)
                 \left( \overline{\mu} \cdot \Xi^{(2)} \right) }  
              { \Xi^{(1)}{}^2 \Xi^{(2)}{}^2
               -\left( \Xi^{(1)} \cdot \Xi^{(2)} \right)^2 } ,\\
\beta^{(2)} &\equiv& 
         \frac{ \Xi^{(1)}{}^2 
                \left( \overline{\mu} \cdot \Xi^{(2)} \right)
               - \left( \Xi^{(1)} \cdot \Xi^{(2)} \right)
                 \left( \overline{\mu} \cdot \Xi^{(1)} \right) }  
              { \Xi^{(1)}{}^2 \Xi^{(2)}{}^2
               -\left( \Xi^{(1)} \cdot \Xi^{(2)} \right)^2 } .
\eea
Then it is easy to show that  $\tilde{\mu}_m$ are invariant under
the transformation $\overline{\mu}_m \rightarrow \overline{\mu}_m +
\alpha^{(1)} \Xi_m^{(1)} + \alpha^{(2)} \Xi_m^{(2)}$ for arbitrary
constants $\alpha^{(1)}$ and $\alpha^{(2)}$.  One can also check that
the number density is invariant under the permutation $\Xi_m^{(1)}
\rightarrow \Xi_m^{(2)} \rightarrow \Xi_m^{(3)} \rightarrow
\Xi_m^{(1)}$, as long as the two conditions used in the definition of
$\tilde{\mu}_m$ are independent. The $B-L$ number density is given by
\bea
\label{eq:b-l_pq}
n_{B-L}(t_D) &=& \sum_m \left(B_m-L_m \right)
n_m (t_D)\non\\
&=& - \left(\Delta_B- \Delta_L \right) \Delta_R \frac{D}{C} 
\frac{k {\cal R} m_{3/2} T_D^2}{6} ,
\eea
where we have defined
\bea
C &\equiv& \Delta_L^2 \lkk B^2 R^2 - \left( B \cdot R \right)^2 \rkk
         + \Delta_B^2 \lkk L^2 R^2 - \left( L \cdot R \right)^2 \rkk 
         \non \\ 
         && + \Delta_R^2 L^2 B^2
         - 2 \Delta_R \lkk \Delta_B L^2 \left( B \cdot R \right)
                         + \Delta_L B^2 \left( L \cdot R \right) \rkk
         \non \\
         &&
         + 2 \Delta_L \Delta_B \left( L \cdot R \right) \left( B \cdot R \right)      
        , \non \\
D &\equiv& L^2 B^2 R^2 - L^2 \left( B \cdot R \right)^2 
                       - B^2 \left( L \cdot R \right)^2  
\eea
Thus,  $B-L$ asymmetry is generated if and only if the interaction
in thermal equilibrium violates both $B-L$ and $PQ$ symmetries, {\it
  i.e.}, $\Delta_B-\Delta_L \ne0$ and $\Delta_R \ne 0$, the meanings
of which are the same as before.

As an illustration we take the operator in Eq.~(\ref{eq:dim5}),
leading to $\Delta_R = 2 ( R_L + R_{H_u})$ and $\Delta_B-
\Delta_L=-2$.
Then the baryon-to-entropy ratio is given by
\bea
\frac{n_B}{s} &=&\frac{120 k {\cal R} ( R_L + R_{H_u})}{
23 \pi^2 g_*} \frac{D}{C}\frac{m_{3/2}}{T_D}\non\\
&\sim& 3 \times 10^{-10} \left(\frac{m_{3/2}}{3 {\rm TeV}}\right) 
\left(\frac{T_D}{10^{12} {\rm GeV}}\right)^{-1},
\eea
where we have assumed that the $PQ$ charges are of the order of
unity.

It should be noted that the constraint on the reheating temperature
due to the gravitino problem is avoided, since the axino,
the superpartner of the axion, naturally exists in our scenario.

\subsection{Large lepton asymmetry}

As a final application we consider a scenario in which a large lepton
asymmetry of electron type and a small baryon asymmetry are generated
simultaneously. We assume that the system has only baryon and lepton
symmetries throughout this subsection for simplicity. With an additional
$U(1)$ symmetry, the main points of the following discussion are
unchanged. In order to generate a positive and large lepton asymmetry of
electron type, we choose a leptonic flat direction such as ``$e^c_{2}
L_{3} L_{1}$'', in which the subscripts denote the generation. We also
assume that  nonrenormalizable operators in the superpotential are
forbidden by some R-symmetry.  Then the NG boson $a_L$ becomes massive
due to the A-term, while $a_B$ remains massless.  Our strategy is as
follows.  In the first place, large lepton asymmetry is generated
through the AD mechanism. Since the large lepton asymmetry is stored in
the AD condensate until it decays at the decay temperature $T_\phi \sim
10$ GeV, it is protected from the sphaleron effects. On the other hand,
a small baryon asymmetry is generated through spontaneous baryogenesis
with a baryon and lepton violating interaction.
It should be noted that the signs of these asymmetries are as
expected, that is, $\dot{a}_L < 0$ induces a positive lepton asymmetry
of electron type in the AD sector, while it leads to a positive baryon
asymmetry if $\Delta_B \Delta_{L} > 0$. Thus we can naturally explain
a positive large lepton asymmetry of electron type and a positive
small baryon asymmetry at the same time.

First we evaluate the large lepton asymmetry in the AD sector.  Since
the AD field is likely to dominate the universe after it starts
oscillating, the resultant lepton asymmetry of electron type is
\beq
\frac{n_L}{s} = \frac{3}{4} \frac{T_\phi}{m_\phi} \lesssim O(1).
\label{eq:ADl}
\eeq

%
When the AD field sits far away from the origin due to the Hubble-induced
mass term Eq.~(\ref{eq:hmass}), $a_L$ slow-rolls as
\beq
\dot{a}_L \simeq  - \frac{k {\cal Q} m_{3/2}^2\phi_D^n}{H_D v_a M_*^{n-2}},
\eeq
where we used Eq.~(\ref{eq:KA}),  the subscript `D' means that the
variable is evaluated at decoupling, and ${\cal Q} = \sum_i L_i = 1$.
Assuming that the symmetry-breaking interaction decouples before 
the AD field starts to oscillate,
the baryon number density is given by
\bea
n_{B}(t_D)& \sim& -
 \Delta_B  \Delta_L          
                  \frac{B^2 L^2}{                     
                        B^2 \Delta_L^2 
                   + L^2 \Delta_B^2                 
                     } 
                     \frac{\dot{a}_L T_D^2}{6 v_a} \non\\
                    &\sim&  \frac{|\dot{a}_L| T_D^2}{ v_a},
\eea
where we have assumed $\Delta_B \Delta_L > 0$ in the second line.
Hence the resultant baryon asymmetry is 
\begin{eqnarray}
\frac{n_B}{s} &=&
\frac{n_{B}(t_D) \left(\frac{m_\phi}{H_D}\right)^{\frac{3}{2}}}{
\frac{1}{2} m_\phi \phi_{osc}^2} \frac{3 T_\phi}{4 m_\phi} \non\\
  &\sim& 10^{-10} \left(\frac{m_{3/2}}{1{\rm TeV}}\right)^2
                    \left(\frac{m_{\phi}}{100{\rm GeV}}\right)^{-\frac{1}{2}}
                          \left(\frac{T_\phi}{10{\rm GeV}}\right)\non\\
                                &&\times
                     \left(\frac{T_D}{10^{10}{\rm GeV}}\right)^{-3}
                                                  \left(\frac{M_{*}}{10^{16}{\rm GeV}}\right)^{-2}
                          \left(\frac{\langle \phi_D \rangle}{M_*}\right)^{(n-2)}\non\\
                          &&\times
                          \left(\frac{\phi_{osc}}{M_*}\right)^{-2},
\end{eqnarray}
where we assumed that the universe is radiation dominated when the
baryon and lepton violating interaction decouples. Although such an
assumption requires a very high reheating temperature, which might be
constrained by the gravitino problem, the entropy production due to
the decay of the AD field weakens the constraint considerably, so we
can safely adopt the assumption.

Recently, it was pointed out that complete or partial equilibrium
between all active neutrinos may be accomplished before BBN through
neutrino oscillations in the presence of neutrino chemical potentials
\cite{equilibrium}. However, even if complete equilibrium is realized,
positive lepton asymmetry of electron type still survives, contrary to
the scenario proposed in \cite{KTY} because the total lepton asymmetry
is positive in this case.

Finally we comment on $Q$-balls. For general flat directions that
include squarks, the numerical coefficient of one-loop corrections,
$K$, is considered to be negative. It is known that the scalar field,
which oscillates in the potential as Eq.~(\ref{eq:grav}) with negative
$K$, experiences spatial instabilities, and deforms into
non-topological solitons, $Q$-balls. However, $K$ can be positive if
the $e_cLL$ direction includes the third generation and $\tan \beta$
is large, so here we have assumed this is the case. Thus we do not
have to take $Q$-balls  into account.

\section{Discussion and conclusions}
\label{sec:con}

In this paper, we have discussed a spontaneous baryogenesis mechanism in
flat directions. First of all, we have identified the Nambu-Goldstone
mode, which derivatively couples to the associated $U$(1) current and
rotates due to the A-term. Such a derivative coupling and a rotation of
the NG mode naturally realize spontaneous baryogenesis in the context
of the flat direction if a current violating interaction exists. We gave
a generic formula for the baryon asymmetry produced.

As concrete examples, we have investigated two scenarios of
baryogenesis in detail. First of all, we considered spontaneous
baryogenesis in a flat direction with vanishing $B-L$ charge. For such
a flat direction, the AD mechanism does not work without $Q$-balls. On
the other hand, we have shown that baryogenesis in such a flat
direction can be easily realized in the context of spontaneous
baryogenesis. All we need is an interaction violating the $B-L$
symmetry.  Such an interaction is, for example, given by the dimension
five operator, which gives the Majorana masses of neutrinos. In
particular, we have discussed spontaneous baryogenesis in a flat
direction with neither baryon nor lepton charge, which was recently
proposed by Chiba and the present authors. It is shown that
baryogenesis is possible if we introduce another global symmetry such
as the $PQ$ symmetry.  Finally, we discussed a scenario in which a
positive and large lepton asymmetry of electron type is compatible
with a positive and small baryon asymmetry. It is shown that it is
possible to realize such a scenario and thereby remove any discrepancy
of baryon asymmetry between those derived from BBN and the CMB.

\subsection*{ACKNOWLEDGMENTS}
 We thank A.D. Dolgov for useful discussions and W. Kelly
  for fruitful comments. M.Y. is grateful to C. Burgess, G.
  Moore, and A. Mazumdar for useful discussions and thanks McGill
  University for hospitality. This work was partially supported by the
  JSPS Grant-in-Aid for Scientific Research No.\ 10975 (F.T.) and
  Research Abroad (M.Y.). M.Y. is partially supported by the
  Department of Energy under Grant No. DEFG0291ER40688.

\end{document}